\documentclass[12pt]{article}
\begin{document}
%\newcounter{popnr}
%\def\fn#1{{\mathop{{\rm #1}}}}
%\def\theequation{\thesection.\arabic{equation}}
%\renewcommand{\theequation}{\arabic{section}.\arabic{equation}}
%\newcommand{\alpheqn}{\setcounter{popnr}{\value{equation}}
%                      \addtocounter{popnr}{1}
%                      \setcounter{equation}{0}
%   
%\renewcommand{\theequation}{\arabic{section}.\arabic{popnr}\alph{e
%quation}}}
%\newcommand{\reseteqn}{\setcounter{equation}{\value{popnr}}
%     \renewcommand{\theequation}
%     {\arabic{section}.\arabic{equation}}}
\def\be{\begin{equation}}
\def\ee{\end{equation}}
\def\bq{\begin{equation}}
\def\eq{\end{equation}}
\def\bqa{\begin{eqnarray}}
\def\eqa{\end{eqnarray}}
\def\roughly#1{\mathrel{\raise.3ex
\hbox{$#1$\kern-.75em\lower1ex\hbox{$\sim$}}}}
\def\lsim{\roughly<}
\def\gsim{\roughly>}
\def\llgm{\left\lgroup\matrix}
\def\rrgm{\right\rgroup}
\def\vectrl #1{\buildrel\leftrightarrow \over #1}
\def\partrl{\vectrl{\partial}}
\def\gslash#1{\slash\hspace*{-0.20cm}#1}

\begin{center}
{\bf\Large Anisotropic geodesic fluid in non-comoving spherical coordinates} 
\end{center}

 \vspace {0.5 cm}
\begin{center}
{\large P.C. Stichel} \\[2.5mm]
{\large Fakult\"{a}t f\"{u}r Physik, Universit\"{a}t Bielefeld }\\[1.2mm] 
{\large D-33501 Bielefeld, Germany} \\
{\large email: peter@physik.uni-bielefeld.de}     \\
\end{center}

\vspace{2 cm}

\baselineskip 20pt

\begin{center}
{\bf Abstract}
\end{center}
We start with a recently introduced spherically symmetric geodesic fluid 
model (arXiv: 1601.07030) whose energy-momentum tensor in the comoving frame 
is dust-like with nontrivial energy flux. In the non-comoving energy frame 
(vanishing energy flux) the same EMT contains besides dust only radial 
pressure. We present Einstein's equations together with the matter equations 
in static spherically symmetric coordinates. These equations are 
self-contained (four equations for four unknowns). We solve them 
analytically except for a resulting nonlinear ordinary differential equation 
(ODE) for the gravitational potential. This ODE can be rewritten as a 
Lienard differential equation which, however, may be transformed into a 
rational Abel differential equation of the first kind. Finally we list some 
open mathematical problems and outline possible physical applications 
(galactic halos, dark energy stars) and related open problems.

\section{Introduction}
Analytic solutions of the coupled Einstein-matter equations for the stationary
anisotropic and spherically symmetric case, without supplying any external
input, are rather rare. The only example we know, a conformal flat
generalization of the de Sitter space-time, has been published very recently
([1], section 6). But normally one has to provide some external input. The
generic case has been discussed recently by Herrera, Ospino and Di Prisco
[2]. The authors of [2] provide only three Einstein equations for five unknowns
(energy density, two pressures, two metric functions). So the knowledge of two
solution generating functions is required. 

In the present paper we describe another model for which we derive an analytic
solution except for one remaining ordinary differential equation.  
We start with a recently introduced irrotational geodesic fluid model whose 
energy-momentum tensor (EMT) in the frame comoving with the fluid is 
dust-like with nontrivial energy flux [3]. Then we pass over to the 
non-comoving energy frame (vanishing energy flux [4]). Here the same EMT 
contains besides dust only radial pressure. We consider the resulting 
Einstein's field equations together with the matter equations in static 
spherically symmetric coordinates. These equations are self-contained 
(four equations for four unknowns). We solve them analytically except for a 
resulting nonlinear ordinary differential equation (ODE) for the 
gravitational potential. This ODE turns out to be the general relativistic 
generalization of a corresponding ODE derived in [5] for the nonrelativistic 
darkon fluid model. It has been used in [5] as a model for galactic halos.

The paper is organized as follows:  We define our model in section 2.  
In section 3 we introduce non-comoving coordinates and present the 
corresponding Einstein equations and the matter equations. In the course 
of integration of these equations in section 4 we derive a nonlinear ODE 
for the gravitational potential. We reformulate this ODE in section 5 as a 
Lienard differential equation and transform it into a rational Abel 
differential equation of the first kind. In section 6 we list some open 
mathematical problems. Finally, possible physical applications 
(galactic halos, dark energy stars) and related open problems are outlined 
in section 7.

\section{Fluid model}

Our model is defined by a self-gravitating, irrotational, pressure-less 
and stress free geodesic fluid whose EMT in the frame comoving with the fluid 
is dust-like with a nontrivial energy flux.

Therefore our model will be described by the following covariant set of 
equations (Greek indices run from 0 to 3 and we use the usual summation 
convention)
\begin{itemize}
\item 
Einstein's equations  ($\kappa = 8 \pi G, ~c=1$)
\begin{equation}
G^{\mu\nu} = \kappa T^{\mu\nu} 
%\label{(1)}
\end{equation}
with a EMT~ $T^{\mu\nu}$, decomposed w.r.t. the unit and time-like fluid 
velocity vector $u^\mu$
\begin{equation}
T^{\mu\nu} = \rho u^\mu u^\nu + q^\mu u^\nu + q^\nu u^\mu 
%\label{(2)}
\end{equation}
where $\rho$ is the total energy density (comprising baryonic and the so 
called dark sector contributions) and $q^\mu$ is the energy flux vector \break
($u^\mu q_\mu = 0$) in the comoving frame.

\break

\item
Constraints for $u^\mu$ 

Geodesic flow:
\begin{equation}
u^\lambda \bigtriangledown_\lambda u^\mu = 0
\end{equation}
Irrotational flow:    
\begin{equation}
\bigtriangledown_\mu u_\nu - \bigtriangledown_\nu u_\mu = 0
\end{equation}
The covariant derivative $\bigtriangledown_\mu$ is given in terms of a 
torsion-free connection (Christoffel symbols).

\item
Covariant conservation of the EMT
\begin{equation}
\bigtriangledown_\mu T^{\mu\nu} = 0
\end{equation}
which is a consequence of (1) (Bianchi identities).

\end{itemize}

\section{Einstein's equations and the matter equations in non-comoving, 
static and spherically symmetric coordinates}

Our fluid will be assumed to move with radial velocity $v$ relative 
to the energy frame (EF). Such a choice of relative motion may be related to 
the observed motion of e.g. a galaxy relative to the microwave 
background [6]. 

For static spherically symmetric coordinates in the EF we use Schwarzschild 
(canonical) coordinates [7]
\begin{equation}
ds^2 = - e^{2\phi (r)} dt^2 + e^{2\lambda(r)} dr^2 + r^2 d\Omega^2
\end{equation}                                 
Then $u^\mu$ and $q^\mu$ are given by $(v = v (r), ~q = q (r))$
\begin{equation}
u^\mu = \beta (n^\mu + v s^\mu), ~~ q^\mu = q \beta (v n^\mu + s^\mu)
\end{equation}
with $\beta = (1 - v^2)^{- 1/2}$. 
The time-like and space-like unit vectors $n^\mu$ and $s^\mu$ are 
defined by
\begin{equation}
n^\mu = \left(e^{-\phi}, 0\right), ~~s^\mu = \left(0, e^{-\lambda}\right).
\end{equation}                                                              
Sometimes it is convenient to use instead of $\lambda(r)$ the mass function
$M (r)$ related to each other by
\begin{equation}
e^{2\lambda} = \left( 1 - \frac{2M}{r}\right)^{-1}
\end{equation}
The EMT (2) reads in the energy frame (decomposition of $T^{\mu\nu}$ 
w.r.t. $n^\mu$ and $s^\mu$)[8]
\begin{equation}
T^{\mu\nu} = \rho^* n^\mu n^\nu + p^*_r s^\mu s^\nu
\end{equation}
where the energy density $\rho^*$ and the radial pressure $p^*_r$
in the EF are related to the corresponding kinematic quantities in the 
comoving frame by
\begin{equation}
\rho^* = \frac{\rho}{1 + v^2},~~~ p^*_r = - \rho^* v^2
\end{equation}
In (11) we have used the relation
\begin{equation}
q = - \rho^* v
\end{equation}
which follows from the requirement of the vanishing energy flux in the EF [8].

With the metric (6) and the EMT (10) we obtain for the Einstein equations 
(1) [7]
\begin{equation}
\kappa \rho^* = \frac{2M^\prime}{r^2}
\end{equation}
\begin{equation}
\kappa ~ p_r^* = \frac{1}{r^2} \left( -1 + \left( 1 - \frac{2M}{r} \right) (1 + 2r
  \phi^\prime) \right)
\end{equation}
\begin{equation}
0 = \left( \phi^{\prime\prime} + \phi^{\prime 2} + \frac{\phi^\prime}{r}\right) \left( 1 -
  \frac{2M}{r} \right) - \left( \phi^\prime + \frac{1}{r}\right) \left(
  \frac{M}{r}\right)^\prime
\end{equation}
                                                                                                                                                                                                                                                                                                                                                                                                                                                                                                  
Here and in what follows a prime denotes differentiation of a function w.r.t. 
to its argument.

\break

The matter equations consist of two parts:
\begin{itemize}
\item 
The generalized Tolman-Oppenheimer-Volkoff (TOV) equation, following 
from the space-like part of (5) (or, more directly, from  (13)-(15))
\begin{equation}
(\rho^* + p^*_r) \phi^\prime + \frac{1}{r^2} (r^2 p^*_r )^\prime = 0
\end{equation}
               
\item
The geodesic flow constraint (3)[8](cp. also [9])
\begin{equation}
v v^\prime \beta^2 + \phi^\prime + 0
\end{equation}
\end{itemize}
                                                                                                                
The irrotational flow constraint (4) is automatically satisfied in 
spherically symmetric coordinates.

\section{Integration of the Einstein and matter equations}

To integrate the set of independent equations (14)-(17) we will proceed in 
three steps:
\begin{itemize}
\item 
Equ. (17) can easily be integrated
\begin{equation}
v^2 = 1 - e^{2\phi}
\end{equation}                                               
                                                                                                                                                                                                                                                                                                                                                                                                                                                                                                                                                                                                                                          
\item        
Insertion of $\rho^* = - p^*_r/v^2$ from (11) and (18) into the
TOV-equation (16) leads to
\begin{equation}
(r^2 p_r^*)~ \frac{e^{2\phi} \phi^\prime}{e^{2\phi} - 1} + (r^2 p_r^*)^\prime 
= 0
\end{equation}
which again can easily be integrated
\begin{equation}
p_r^* = \frac{\alpha}{r^2} \left( 1 - e^{2\phi}\right)^{-1/2}
\end{equation}
where $\alpha$ is an integration constant.

For $ p_r^*$ to be real valued we have to require
\begin{equation}
\phi (r) \le 0
\end{equation} 
              
\item
Next we insert (20) into the 2nd Einstein equation (14) and solve for $M/r$
\begin{equation}
1 - \frac{2M}{r} = \frac{\kappa\alpha (1 - e^{2\phi} )^{-1/2} + 1}{1 + 2r 
\phi^\prime}
\end{equation}                                                                                                                
Insertion of (22) into the 3rd Einstein equation (15) leads, after some 
straightforward manipulations, to a nonlinear ordinary differential equation 
(ODE) for the gravitational potential
\begin{eqnarray}
&& \left( (r^2 \phi^\prime)^\prime + 2 r^2 \phi^{\prime 2} \right) \left( (1 -
  e^{2\phi} )^{3/2} + \kappa\alpha ( 1 - e^{2\phi} ) \right) \nonumber \\
&& + \frac{\kappa\alpha}{2} ( 1 + r\phi^\prime ) (1 + 2r \phi^\prime) e^{2\phi}
  = 0
\end{eqnarray}

\end{itemize}
{\bf Comment:} The weak field limit ($\phi = 0 (\epsilon)$) of (23) yields 
(cp. [8], subsection 6.5)
\begin{equation}
( r^2 \phi^\prime)^\prime = \frac{\gamma}{2} ( - 2\phi)^{-3/2}
\end{equation}
where we have put $\gamma = - \kappa\alpha,~ (\gamma = 0 (\epsilon^{5/2}))$.

The result (24) turns out to be equal to the corresponding equation obtained
for the stationary solution of the nonrelativistic darkon fluid model in
[5]. In this limit we get from the Poisson equation and (24) the following
relation between the energy density $\rho$ and the potential $\phi$ 
\begin{equation}
\kappa \rho = \frac{\gamma}{r^2} ( - 2 \phi)^{-3/2}
\end{equation}
Then positivity of $\rho$ requires
\begin{equation}
\gamma > 0
\end{equation}
Note that a positive energy density yields a negative radial pressure according
to (11).

\section{Reformulation of the ODE (23) as a Lie-nard differential equation or 
as an Abel differential equation}\footnote{For convenience, by ``Abel 
differential equation'' we mean Abel differential equation 
of the first kind}

With the transformation
\begin{equation}
r \rightarrow x = \log r ,~~~~~ \phi (r) \rightarrow \varphi (x) = - 2 \phi (r)
\end{equation}
the ODE (23) becomes the autonomous ODE
\begin{equation}
(\varphi^{\prime 2} - \varphi^{\prime\prime} - \varphi^\prime) ( ( 1 -
e^{-\varphi})^{3/2} - \gamma (1 - e^{-\varphi})) 
= \gamma e^{-\varphi} (1 - \varphi^\prime ) ( 1 - \varphi^\prime / 2 ).
\end{equation} 
The further transformation
\begin{equation}
\varphi \rightarrow f = 1 - e^{-\varphi}
\end{equation}
leads to the mixed Lienard differential equation
\begin{equation}
f^{\prime\prime} + g_1 (f) f^\prime + g_2 (f) f^{\prime 2} + g_0 (f) = 0
\end{equation} 
with
\begin{eqnarray}
g_0 (f) & = & \gamma (1 - f)^2 (f^{3/2} - \gamma f)^{-1} \nonumber \\
g_1 (f) & = & \left( f^{3/2} + \gamma \frac{f}{2} - \frac{3}{2} \gamma \right) 
( f^{3/2} - \gamma f)^{-1} \nonumber \\
g_2 (f) & = & \frac{\gamma}{2} ( f^{3/2} - \gamma f )^{-1}  
\end{eqnarray}                                           
To transform (30) into an Abel differential equation we proceed as usual [10]:

\noindent
With $f (x) \rightarrow y (f)$, 
\begin{equation}
y (f) = (1 - \gamma f^{-1/2} )^{-1} ( f^\prime (x))^{-1}
\end{equation}
we obtain from (30)
\begin{equation}
y^\prime (f) = h_2 (f) y^2 (f) + h_3 (f) y^3 (f)
\end{equation}
with
\begin{eqnarray}
h_2 (f) & = & 1 + \frac{\gamma}{2} f^{-1/2} - \frac{3}{2} \gamma f^{-3/2}
\nonumber \\
h_3 (f) & = & \gamma f^{-3/2} (1 - f)^2 (1 - \gamma f^{-1/2}). 
\end{eqnarray}
By the further transformation
\begin{equation}
f \rightarrow x = f^{1/2},~~~ u (x) = y (f)
\end{equation}
we obtain from (33),(34) a rational Abel differential equation 
\begin{equation}
u^\prime (x) = H_2 (x) u^2 + H_3 (x) u^3
\end{equation}
with 
\begin{eqnarray}
H_2 & = & \frac{2 x^3 + \gamma x^2 - 3\gamma}{x^2} \nonumber \\
H_3 (x) & = & \frac{2\gamma (x -\gamma) (1 - x^2)^2}{x^3}
\end{eqnarray}
Unfortunately (36) does not belong to the known integrable cases of rational 
Abel differential equations. But, as shown in [11](see also [12] and [13]), 
all integrable rational Abel differential equations consist of classes whose
members are related to each other by the equivalence transformation
\begin{equation}
x = F (z),~~ u(x) = P (z) w (z) + Q (z)
\end{equation}
where F, P and Q are arbitrary functions of $z$ satisfying                     
$F^\prime \not= 0$ and $P \not= 0$.
                    
A computer algebra routine has been presented in [11] which allow us to 
decide whether a given Abel differential equation belongs to one of the known 
integrable classes.

\section{Open problems}

From the results of section 5 follow immediately the following open 
mathematical problems: 
\begin{itemize}
\item
For which values of $\gamma$ does the Lienard equation (30) has positive 
solutions f(x) with $0 \le f (x) \le 1$
\item
Check by means of the computer program presented in [11] whether the Abel 
equation (36) belongs to one of the known integrable classes.
\item
If the answer is no, elaborate numerical solutions for equation (36).
\item
Stability of the stationary solutions.
\end{itemize}                                                                                                                                                                                                   

\section{Physics}
Analytic or numerical results for the gravitational potential $\phi (r)$
from solutions of either the nonlinear ODE (23) or of any of its 
equivalent forms given in section 5 will be suitable for the description of 
either galactic halos or of dark energy stars.

\subsection{Galactic halos}
A star in circular motion in a gravitational potential $\phi (r)$ possesses 
the tangential velocity $ v_{tg} (r)$ given by the relationship
\begin{equation}
v^2_{tg} = r \phi^\prime
\end{equation}
which holds also in the general relativistic case (see [14]). 
Keeping in mind that in our model $\phi$ is sourced not only by stellar 
matter but also by the so called dark sector contributions, we may use a 
solution for $\phi$ in (39) for modeling of galactic rotation curves (RCs).

For the description of galactic halos we need gravitational potentials $\phi$ 
which vanish for $r \rightarrow\infty$. But in the nonrelativistic case, 
described by (24), solutions vanish already for a finite but very large 
distance as shown in [5] by a theorem due to Taliaferro [15] as well as by 
numerical results.

{\bf Problem 1:} Will admissible solutions (see section 6) of the Lienard 
equation (30) extent up to $x \rightarrow + \infty$             
or will they end at finite $x$?

The numerical results for the RCs in the nonrelativistic case as shown in 
([5], Fig.6) for our model are in good agreement with the observed nearly 
flat RCs at large radii for ``dark matter dominated''  galaxies (for a very 
recent review on the dark matter issues see [16]). 

We do not expect any essential modifications of the weak field limit at large 
radii for the relativistic model presented in this paper.

On the other hand, at small scales, our nonrelativistic model seems to show 
numerically a vanishing RC already at a very small but nonzero radius [5]. 
It was not possible to give a definite answer to this point in [5] because 
the numerical solutions have shown a discontinuous behavior around the 
critical value $b_c$ for $b = \varphi^\prime(0)$.   
                          
{\bf Problem 2:} 
Behavior of the gravitational potential $\phi (r)$ for $r\rightarrow 0$ 
in the weak field limit as well as for the case of strong fields.

\subsection{Dark energy stars}
As has been already stated, our model shows a negative radial pressure for 
a positive energy density. Therefore it is predestinated for the description 
of anisotropic dark energy (DE) stars. To do that we have to take a solution 
of our model as interior solution which have to be matched with the exterior 
Schwarzschild solution.

Recent treatments of anisotropic DE stars are to be found in [17], [18] and 
[19]. In all these cases analytic solutions are given by supplying some 
functions and constants as external input. In [17] proportionalities between 
energy densities and DE radial pressure as well as analytic expressions for 
the two metric functions $\phi$ and $\lambda$ are assumed. In [18] two 
equations of state and an analytic expression for the DE energy density are 
given. An analytic expression for the mass function and a DE- equation of 
state for the radial pressure are provided in [19].

In our model no such external inputs are needed. But in order to proceed we 
have to succeed in finding (approximate) analytic expressions or numerical 
results for the gravitational potential.


\begin{thebibliography}{99}

\bibitem{[1]}\O . Gr\o n and S. Johannesen, Conformally flat spherically symmetric
  spacetimes, 
Eur. Phys. J. Plus (2013), 128:92

\bibitem{[2]}L. Herrera et al., All static spherically symmetric anisotropic 
solutions of Einstein's equations, 
      Phys. Rev. D 77, 027502 (2008); arXiv: 0712.0713

\bibitem{[3]} P. Stichel, Cosmological model with dynamical curvature, 
arXiv: 1601.07030

\bibitem{[4]} M. Bruni et al., COSMOLOGICAL PERTURBATIONS AND THE PHYSICAL 
MEANING OF GAUGE-INVARIANT VARIABLES, Astrophys. J. 395, 34 (1992)

\bibitem{[5]} P. Stichel and W. Zakrzewski, Nonstandard Approach to Gravity 
for the Dark Sector of the Universe, Entropy, 15, 559 (2013)

\bibitem{[6]} L. Herrera, ON THE MEANING OF GENERAL COVARIANCE AND THE 
RELEVANCE  OF OBSERVERS IN GENERAL RELATIVITY, Int. J. Mod. Phys. D 20, 
2773 (2011) 

\bibitem{[7]} H. Stephani et al., Exact Solutions of Einstein's Field 
Equations, Cambridge University Press, 2006

\bibitem{[8]} P. Stichel and W.Zakrzewski, Eur. Phys. J. C (2015), 75:9; 
arXiv: 1409.1336

\bibitem{[9]} L. Herrera et al., Spherically symmetric dissipative anisotropic 
fluids: A general study, 
         Phys. Rev. D 69, 084026 (2004)

\bibitem{[10]} A.D.  Polyanin and V.F. Zaitsev, EXACT SOLUTIONS FOR ORDINARY 
DIFFERENTIAL EQUATIONS, 
       Chapman \& Hall/CRC, 2003

\bibitem{[11]} E.S. Cheb-Terrab and A.D. Roche, Abel ODEs: Equivalence and 
Integrable Classes, 
        Comput. Phys. Commun. 130(1-2), 204 (2000); arXiv: math-ph/ 0001037

\bibitem{[12]} E.S. Cheb-Terrab and A.D. Roche, An Abel ordinary differential 
equation class generalizing known 
        integrable classes, Eur J Appl Math 14(2), 217 (2003); 
arXiv: math/0002059

\bibitem{[13]} J. Gine and J. Llibre, On the integrable rational Abel 
differential equations, ZAMP 61, 33 (2010) 

\bibitem{[14]} S. Capozziello et al., Hybrid metric-Palatini gravity, 
Universe 1(2), 199 (2015); arXiv: 1508.04641

\bibitem{[15]} S. Taliaferro, On the positive solutions of $y^{\prime\prime} +
  \phi (t) y^{-\lambda} = 0$, Nonlin. Anal.Theor. 2, 437 
        (1978)

\bibitem{[16]} S. Tulin and Hai-Bo Yu, Dark Matter Self-interactions and 
Small Scale Structure,
         arXiv: 1705.02358

\bibitem{[17]} A. Yadav et al., Singularity-free dark energy star, Gen. Rel. 
Grav. 44, 107 (2012); 
        arXiv: 1102.1382

\bibitem{[18]} M. Kalam et al., Anisotropic Quintessence stars, arXiv: 1308.
0147

\bibitem{[19]} P. Bhar et al., Dark energy stars: Stable configurations, 
arXiv: 1610.01201

\end{thebibliography}
\end{document}